\documentclass[twocolumn,amsmath,amssymb,floatfix,nofootinbib]{revtex4}
\usepackage{epsfig}
\usepackage{amsmath,amssymb}
\usepackage[all]{xy}

\usepackage{graphicx}
\usepackage{epstopdf}
\newcommand{\lsi}{\,\raisebox{-0.13cm}{$\stackrel{\textstyle<}
{\textstyle\sim}$}\,}
\newcommand{\gsi}{\,\raisebox{-0.13cm}{$\stackrel{\textstyle>}
{\textstyle\sim}$}\,}

\begin{document}

\title{Dark Matter and the Baryon Asymmetry of the Universe}

\author{Glennys R. Farrar and Gabrijela Zaharijas}

\affiliation{Center for Cosmology and Particle Physics,
Department of Physics\\ New York University, NY, NY 10003,USA\\
}

\begin{abstract}{We present a mechanism to generate the baryon asymmetry of the Universe which preserves the net baryon number created in the Big Bang.  If dark matter particles carry baryon number $B_X$, and  
$\sigma^{\rm ann}_{\bar{X}} < \sigma^{\rm ann}_{X } $, the $\bar{ X}$'s freeze out at a higher temperature and
have a larger relic density than $X$'s.   If $m_X \lsi  4.5 \,B_X \,$GeV and the annihilation cross sections differ by $\mathcal{O}$(10\%) or more,  this type of scenario naturally explains the observed $\Omega_{DM} \approx 5\, \Omega_b$.   
}
\end{abstract}
 \maketitle

The abundance of baryons and dark matter (DM) in our Universe poses
several challenging puzzles:\\
$\bullet$ Why is there a non-zero net nucleon density and what
determines its value?\\ 
$\bullet$ What does dark matter consist of? \\
$\bullet$  Is it an accident that the dark matter density is roughly comparable
to the nucleon density, $\rho_{DM} = 5 ~\rho_N$? 

As pointed out by Sakharov\cite{sakharov}, baryogensis requires three conditions: non-conservation of baryon number,  violation of C and CP, and a departure from thermal equilibrium.  The last is provided by the expansion of the Universe and the first two are naturally present in unified theories and even in the standard model at temperatures above the electroweak  phase transition.  However in most approaches the origins of DM and the BAU are completely unrelated and their densities could naturally differ by many orders of magnitude.  In this Letter we propose a new type of scenario, in which the observed baryon asymmetry is due to the separation of baryon number between ordinary matter and dark matter and not to a net change in the total baryon number since the Big Bang. (See \cite{other} for other papers with this aim.)  Thus the abundances of nucleons and dark matter are related.  The first Sakharov condition is not required, while the last two remain essential.  We give explicit examples in which  anti-baryon number is sequestered at temperatures of order 100 MeV.  

The CPT theorem requires that the total interaction rate of any ensemble of particles and antiparticles is the same as for the  conjugate state in which each particle is replaced by its antiparticle and all spins are reversed.  However individual channels need not have the same rate so, when CP is violated, the annihilation rates of the CP reversed systems are not in general equal.  A difference in the annihilation cross section, $\sigma^{\rm ann}_{\bar{X}} <  \sigma^{\rm ann}_{X } $, means that the freeze out temperature for $X$'s ($T_X$) is lower than for $\bar{X}$'s ($T_{\bar{X}}$).   After the $\bar{X}$'s freeze out, the $X$'s continue to annihilate until the temperature drops to $T_{X}$, removing $B_X$ antinucleons for each $X$ which annihilates.  

Assuming there are no other significant contributions to the DM density, the present values $n_{o\, N}$, $n_{o\, X}$ 
and $n_{o\, \bar{X}}$  are
determined in terms of $m_X$, $B_X$ and the observables $ \frac{\Omega_{DM}}{\Omega_b}$ and $\frac{n_{o\,N}}{n_{o\, \gamma}} \equiv \eta_{10} \,10^{-10}$ or $\rho_{\rm crit}$.  (Following common usage, the subscript $b$ refers to the baryon number in ordinary matter.)  From WMAP, $\eta_{10} = 6.5^{+0.4}_{-0.3}$, $\Omega_m h^2 = 0.14 \pm 0.02$, $\Omega_b h^2 = 0.024 \pm 0.001$\cite{WMAP},  so $\frac{\Omega_{DM}}{\Omega_b} = 4.83 \pm 0.87$. 
Given the values of these observables, we can ``reverse engineer" the process of baryon-number-segregation.  

For brevity, suppose there is only one significant species of DM particle.  Let us define $\epsilon = \frac{n_X}{n_{\bar{X}} }$.  Then the total energy density in $X$'s and $\bar{X}$'s is  $\rho_{DM} = m_X n_{\bar{X}} (1 + \epsilon)$.  By hypothesis, the baryon number density in nucleons equals the antibaryon number density in $X $ and $\bar{X}$'s, so  $ B_X n_{\bar{X}} (1-\epsilon) = (n_N - n_{\bar{N}}) = \frac{\rho_b}{m_N}$.  Thus  
\begin{equation} \label{kappa}
\frac{\Omega_{DM}}{\Omega_b} = \left( \frac{1 + \epsilon}{1 - \epsilon} \right) \frac{m_X}{m_N B_X}.
\end{equation}
As long as the DM particle mass is of order hadronic masses and $\epsilon $ is not close to 1, this type of scenario naturally accounts for the fact that the DM and ordinary matter densities are of the same order of magnitude.  Furthermore, since $ \frac{1 + \epsilon}{1 - \epsilon} \ge 1$, the DM density in this scenario must be {\it  greater} than the nucleonic density as observed, unless $m_X < m_N B_X$.  

Given the parameters of our Universe, we can instead write (\ref{kappa}) as an equation for the DM mass
$
 m_X = \left( \frac{1 - \epsilon}{1 + \epsilon} \right) \frac{\Omega_{DM}}{\Omega_b} \, B_X m_N .
$
For low baryon number, $B_X = 1\, (2)$, this implies $m_X \lsi 4.5 \,(9)\,$GeV.  If dark matter has other components in
addition to the $X$ and $\bar{X}$, the $X$ must be lighter still.   The observed BAU can be due to baryon number sequestration with heavy DM only if $B_X$ is very large, e.g., strangelets or Q-balls.  However segregating the baryon number in such cases is challenging.  

As an existence proof and to focus our discussion of the issues, we present two concrete scenarios.  In the first, $X$ is a particle already postulated in QCD, the $H$ dibaryon ({\it uuddss}). New particle physics is necessary, however, because the CP violation of the Standard Model via the CKM matrix cannot produce the required $\mathcal{O}$(20\%) difference in annihilation cross sections, since only the first two generations of quarks are involved.  (Readers interested in why the $H$ may have escaped detection in particle physics searches and why its lifetime could be of order the age of the Universe are refered to \cite{fz:H} and references therein.)  The second scenario postulates a new particle with mass  $\lsi 4.5$ GeV, which couples to quarks through dimension-6 operators coming from beyond-the-standard-model physics.   In this case CP violation is naturally large enough, $\mathcal{O}$(10\%), because all three quark generations are involved and, in addition, the new interactions in general violate CP.   After deducing the properties required of these particles by cosmology, we discuss detection methods and then briefly mention particle physics aspects.  As we shall show, the $H,\,\bar{H}$ scenario can already be ruled out by limits on the heat production in Uranus.  

The annihilation rate of particles of type $j$ with particles of type $i$ is
$\Gamma^{ann}_{j}(T) = \Sigma_i ~n_i(T) <\sigma_{ij}^{ann}
v_{ij}>$, 
where $<...>$ indicates a thermal average and $v_{ij}$ is the
relative velocity.  As the Universe cools, the densities of all the
particle species $i$ decrease and eventually the rate of even the most important
annihilation reaction falls below the expansion rate of the
Universe.  The temperature at which this occurs is called the
freezeout temperature $T_j$ defined by $
\Gamma^{ann}_{j}(T_j) = H(T_j) = 1.66 \sqrt{g_*} ~ T_j^2/ M_{Pl} $, 
where $g_*$ is the effective number of relativistic degrees of freedom\cite{kolbTurner}.  Between a few MeV and the QCD phase transition only neutrinos, $e^\pm$ and $\gamma$'s are in equillibrium and $g_* = 10.75$.  Above the QCD phase transition which is presumably around 150 MeV, light quarks and antiquarks ($q, \, \bar{q}$) and $c, \, \bar{c}$, $\mu^\pm$ are also in equilibrium, giving $g_* = 56.25$.  Above $T_b \approx 170$ MeV, $b$ and $\bar{b}$ quarks are kept in equilibrium through $ q~\bar{q}  \leftrightarrow b ~ \bar{b} $ giving $g_* = 66.75$; below this temperature production cannot keep up with decay.  The equilibrium density at freeze out temperature, $n_j(T_j)$, is a good estimate of the relic abundance of the $j$th species\cite{kolbTurner}.   A key element of baryon-number sequestration is that self-annihilation cannot be important for maintaining equilibrium prior to freeze out.  This is easily satisfied as long as $\sigma_{\bar{X}X}^{\rm ann}$ is not much greater than $\sigma_{\bar{X}N}^{\rm ann}$ , since at freezeout in the $H,\,\bar{H}$ and ``$X_4$" scenarios, $n_{H ,\, \bar{H} } \lsi 10^{-5} n_{N ,\, \bar{N} }$ and $n_{X_4 ,\, \bar{X_4
} }\sim 10^{-11} n_{d ,\, \bar{d} }$.  

Given $m_X,\, B_X$ and $g_X$ (the number of degrees of freedom of the $X$ particle) and associated densities $n_{\{X,\bar{X}\}}$, the temperature $T_{\bar{X}}$ at which $\bar{X}$'s must freeze out of thermal equilibrium satisfies: 
\begin{equation} \label{Xbarfo}
  \frac{n_{\bar{X}} - n_X}{n_{\bar{X}}} \frac{n_{\bar{X}}}{n_\gamma } = (1-\epsilon) \frac{ \pi^2 g_X
  x_{\bar{X}}^{3/2} e^{-x_{\bar{X}} } }{2 \zeta(3) (2 \pi)^{3/2} }= 
  \frac{10.75}{3.91}\frac{\eta_{10} 10^{-10} }{B_X} ,
\end{equation}
where $x_{\bar{X}} \equiv m_X/T_{\bar{X}}$. 
$ \frac{10.75}{3.91}$ is the factor by which $\frac{n_b}{n_\gamma}$ increases above $e^\pm$ annihilation.  The equation for $X$ freezeout is the same, with $(1-\epsilon) \rightarrow (1-\epsilon)/\epsilon $.  Freezeout parameters for our specific models are given in Table I; $\tilde{\sigma} \equiv \langle \sigma^{\rm ann} v \rangle / \langle v \rangle$ is averaged over the relevant distribution of c.m. kinetic energies, thermal at $\approx 100$ MeV for freezeout, or the degraded halo distribution with $v_H \lsi 11$km/s for $H,\,\bar{H}$ DM detection discussed next.

If $X$s interact weakly with nucleons, standard WIMP searches constrain the low energy scattering cross section  $\sigma_{DM} \equiv (\sigma^{\rm el}_{\bar{X} N} + \epsilon \sigma^{\rm el}_{XN})/(1+ \epsilon)$.  However if the $X$ is a hadron, multiple scattering in the earth or atmosphere above the detector can cause a significant fraction to reflect or be degraded to below threshold energy before reaching a deep underground detector.  Scattering also greatly enhances DM capture by Earth, since only a small fraction of the halo velocities are less than $v_{\rm esc}^{E} = 11$ km/s.  Table I gives the total fluxes and the factor $f_{\rm cap}$ by which the flux of captured $\bar{X}$'s is lower, for the two scenarios.  These are the result of integrating the conservative halo velocity distribution\cite{zf:window}.  A comprehensive reanalysis of DM cross section limits including the effect of multiple scattering has recently been given in ref. \cite{zf:window}.  A window in the DM exclusion was discovered for $m_X \lsi 2.4$ GeV and $ \tilde{\sigma}_{DM} \approx 0.3 - 1\, \mu $b; otherwise, if the DM mass $\lsi 5$ GeV, $\tilde{\sigma}_{DM}$ must be $ \lsi 10^{-38} {\rm cm}^2$.  

Since $\sigma_{\{X,\bar{X} \}N}$ is negligible compared to $\sigma_{NN}$ and the $X,\,\bar{X}$ do not bind to nuclei\cite{fz:nucbind}, nucleosynthesis works the same in these scenarios as with standard CDM.  Primordial light element abundances constrain the {\it nucleon} -- not {\it baryon} -- to photon ratio! 
\begin{table}[htb] \label{table}

\caption{Freezeout parameters, solar system average $\bar{X}$ flux, and fraction of flux captured in Earth, in two models. } 

\begin{center}
\begin{tabular}{|c|c|c|c|c|c|c|}

\hline
Model & $T_{\bar{X}}$ MeV & $T_{X}$  &  $\tilde{\sigma}^{\rm ann}_{\bar{X}}$ cm$^2$ & $\tilde{\sigma}^{\rm ann}_{{X}}$ & $\Phi_{\bar{X}} ({\rm cm}^2 s)^{-1}$  &	$f^E_{\rm cap}$  \\  \hline 
H    &      86.3       &   84.5  & 	$2.2~10^{-41}$     &    $2.8~10^{-41}$     &      $3.7 \times 10^5$    &	0.20 \\ \hline
$X_4$    &      180    &   159       &		$3.3~10^{-45}$  &    $3.7~10^{-45}$     &     $ 2.9 \times 10^5$   &  $2 \, 10^{-6}   $     \\ \hline
\end{tabular}
\end{center}
\end{table}

The CPT theorem requires that $\sigma^{\rm ann}_{X} + \sigma^{\rm non-ann}_{X} = \sigma^{\rm ann}_{\bar{X}} + \sigma^{\rm non-ann}_{\bar{X}}$.  Therefore a non-trivial consistency condition in this scenario is $ \sigma^{\rm ann}_{X} - \sigma^{\rm ann}_{\bar{X}} \le \sigma^{\rm non-ann}_{\bar{X}} $.  The value of the LHS needed for B-sequestration from Table I is compatible with the upper limits on the RHS from DM searches, and $\sigma^{\rm non-ann}_{\bar{X}} \ge \sigma^{\rm el}_{\bar{X}} $, so no fine-tuning is required to satisfy CPT.     

B-sequestration scenarios imply the possiblity of detectable annihilation of $\bar{X}$'s with nucleons in the Earth, Sun or galactic center.   The rate of $\bar{X}$ annihilation in an Earth-based detector is the $\bar{X}$ flux at the detector, times  $\sigma_{\bar{X}N}^{\rm ann}$, times (since annihilation is incoherent)  the number of target nucleons in the detector,  $6 \times 10^{32} $ per kton.  The final products of $\bar{X} N$ annihilation are mostly pions and some kaons, with energies of order 0.1 to 1 GeV.  The main background in SuperK at these energies comes from atmospheric neutrino interactions whose level is $\sim100$ events per kton-yr\cite{SKatmneutrinos}.  Taking $\Phi^{SK}_{\bar{H}} \approx f_{\rm cap} \Phi_{\bar{H}}$ and $\Phi^{SK}_{\bar{X_4}} =  \Phi_{\bar{X_4}}$ from Table I, the annihilation rate in SuperK is lower than the background if $\tilde{\sigma}^{\rm ann}_{\bar{H}N} \le 6 \times 10^{-44}\,  {\rm cm}^2$ and $\tilde{\sigma}^{\rm ann}_{\bar{X_{4}}N} \le 2 \times 10^{-44}\,  {\rm cm}^2$.   The total energy release of $m_X + B_X m_N $ should give a dramatic signal, so it should be possible for SuperK to improve this limit.  Note that for the $H,\,\bar{H}$ scenario this limit is already uncomfortable, since it is so much lower than the effective cross section required at freezeout.  However this cannot be regarded as fatal, until one can exclude with certainty the possibility that the annihilation cross section is very strongly energy dependent.

Besides direct observation of annihilation with nucleons in a detector, constraints can be placed from indirect effects of $\bar{X}$ annihilation in concentrations of nucleons.  We first discuss the photons and neutrinos which are produced by decay of annihilation products.  The signal is proportional to the number of nucleons divided by the square of the distance to the source, so Earth is a thousand-fold better source for a neutrino signal than is the Sun, all other things being equal.  Since $\gamma$'s created by annihilation in the Earth or Sun cannot escape, the galactic center is the best source of $\gamma$'s; we do not pursue this here because {\it i)} the constraints above imply the signal is several orders of magnitude below present detector capabilities and {\it ii)} $\sigma^{\rm ann}_{\bar{X_4}N}$ is model dependent and may be very small, as discussed below.   

The rate of observable neutrino interactions in SuperK is 
\begin{equation}  \label{neutintsSK}
\Gamma_{\nu{ \rm SK}} =  N_{{\rm SK}}\, \Sigma_i \int{ \frac{d n_{\nu_i}}{dE} \sigma^{\rm eff}_{\nu_i N}  \Phi_{\nu_i}  dE }, 
\end{equation}
where the sum is over neutrino types, $N_{{\rm SK}}$ is the total number of nucleons in SuperK, $\frac{d n_{\nu_i}}{dE}$ is the spectrum of $i$-type neutrinos from an $\bar{X}$ annihilation, $\sigma^{\rm eff}_{\nu_i N} $ is the neutrino interaction cross section summed over observable final states (weighted by efficiency if computing the rate of observed neutrinos), and $ \Phi_{\nu_i} $ is the $\nu_i$ flux at SK.  This last is $f_{\nu_i}$, the mean effective number of $\nu_i$'s produced in each $\bar{X}$ annihilation discussed below, times the total rate of $\bar{X}$ annihilation in the source, $\Gamma^{\rm ann}_{\bar{X},s}$, divided by $\approx 4 \pi R_s^2$, where $R_s$ is the distance from source to SuperK;  $R_s \approx R_E$ for annihilation in Earth.  

In general, computation of the annihilation rate $\Gamma^{\rm ann}_{\bar{X},s}$ is a complex task because it involves solving the transport equation by which DM is captured at the surface, migrates toward the core and annihilates, eventually evolving to a steady state distribution.  However if  the characteristic time for a DM particle to annihilate, $\tau^{\rm ann} = \,<\sigma^{\rm ann} n_N v>^{-1}$, is short compared to the age of the system, equilibrium between annihilation and capture is established (evaporation can be neglected for $M_{DM} \gsi \mathcal{O}$(GeV)\cite{krauss}) so $\Gamma^{\rm ann}_{\bar{X},E}$ equals $f_{\rm cap}  \Phi_{\bar{X}} 4 \pi R_E^2$.  Then the neutrino flux (\ref{neutintsSK}) is independent of $\sigma^{\rm ann}_{\bar{X}N},$ because the annihilation rate per $\bar{X}$ is proportional to it but the equilibrium number of $\bar{X}$'s in Earth is inversely proportional to it.  For Earth, the equilibrium assumption is applicable for $\tilde{\sigma}^{\rm ann} \gsi 5 \times 10^{-49} {\rm cm}^2$, while for the Sun it is applicable if, roughly, $\tilde{\sigma}^{\rm ann} \gsi 10^{-52} {\rm cm}^2$.  For lower annihilation cross sections, transport must be treated to get an accurate estimate, but the equilibrium rate is an upper limit.  

The final state in $\bar{H} N$ annihilation is expected to contain $\bar{\Lambda}$ or $\bar{\Sigma}$ and a kaon, or $\bar{\Xi}$ and a pion, and perhaps additional pions.  In a dense environment such as the core of the Earth, the antihyperon annihilates with a nucleon, producing pions and at least one kaon.  In a low density environment such as the Sun, the antihyperon decay length is typically much shorter than its interaction length.  In Earth, pions do not contribute significantly to the neutrino flux because $\pi^0$'s decay immediately to photons, and the interaction length of $\pi^\pm$'s is far smaller than their decay length so they eventually convert to $\pi^0$'s through charge exchange reactions;  similarly, the interaction lengths of $K^{0}_L$'s and $K^\pm$'s are much longer than their decay lengths, so through charge exchange they essentially all convert to $K^0_S$'s before decaying.  The branching fraction for production of $\nu_{e,\mu}$ and $\bar{\nu}_{e,\mu}$ from $K_S^0 \rightarrow \pi l^\pm \nu$ is $3.4 \times 10^{-4}$ for each, so $f_{\nu_i} \ge 2(3.4\times 10^{-4})$ for $\bar{H}$ annihilation in Earth.  Since the Sun has a paucity of neutrons, any kaons in the annihilation products are typically $K^+$ and furthermore their charge exchange is suppressed by the absence of neutrons.  The branching fraction for $K^+ \rightarrow \mu^+ \nu_\mu$ is 63\% and the $\nu_\mu$ has 240 MeV if the kaon is at rest.   If the final states of $\bar{H}$ annihilation typically contain kaons, then $f_\nu $ is $\mathcal{O}$(1).  However if annihilation favors $\bar{\Xi}$ production, $f_\nu$ could be as low as $\approx 3 \cdot 10^{-4}$ for production of $\bar{\nu}_{e}$'s and $\bar{\nu}_\mu$'s above the charged current threshold.  Thus the predicted neutrino signal from $\bar{H}$ annihilation in the Sun is far more uncertain than in Earth.  

Neutrinos from $\bar{X}$ annihilation can be detected by SuperK, with a background level and sensitivity which depends strongly on neutrino energy and flavor.  Taking the captured $\bar{X}$ flux on Earth from Table I, assuming the neutrinos have energy in the range 20-80 MeV for which the background rate is at a minimum, and taking the effective cross section with which $\nu$'s from the kaon decays make observable interactions in SuperK to be $10^{-42} {\rm cm}^2$, (\ref{neutintsSK}) leads to a predicted rate of excess events from annihilations in Earth of $ \Gamma_{\nu{ \rm SK}} \approx 2$/(kton yr) in the $\bar{H}$ scenario.  This is to be compared to the observed event rate in this energy range $\approx 3$/(kton yr)\cite{SKloEneutrinos}, showing that SuperK is potentially sensitive.  If a detailed analysis of SuperK's sensitivity were able to establish that the rate is lower than this prediction, it would imply either that the $H, \bar{H}$ model is wrong or that the annihilation cross section is so low that the equilibrium assumption is invalid, i.e., $\sigma^{\rm ann}_{\bar{H}N} \lsi 2 \times 10^{-48} {\rm cm}^2$.  The analogous calculation for the Sun gives $ \Gamma_{\nu{ \rm SK}} \approx 130 f_\nu$/(kton yr) for energies in the sub-GeV atmospheric neutrino sample, for which the rate is $\approx 35$ events/(kton yr)\cite{SKatmneutrinos}.\footnote{This estimate disagrees with that of Goldman and Nussinov (GN)\cite{GN}, independently of the question of the value of $f_\nu$.   GN use an $\bar{H}$ flux in the solar system which is eight times larger than our value in Table I from integrating the normal component of the halo velocity distribution, due to poor approximations and taking a factor-of-two larger value for the local DM density.  We include a factor 0.35 for the loss of $\nu_\mu$'s due to oscillation, we account for the fact that only neutrons in SuperK are targets for CC events, and we avoid order-of-magnitude roundup.  Note that the discussion of the particle physics of the $H$ in \cite{GN} applies to the case of an absolutely stable $H$, which we discussed but discarded in \cite{fz:nucbind,fz:H}.  }  Thus if $f_\nu$ were large enough, SuperK could provide evidence for the $H,\,\bar{H}$ scenario via energetic solar neutrinos.  However the absence of a solar neutrino signal cannot be taken as excluding the $H,\,\bar{H}$ scenario, given the possibility that $f_\nu \le 10^{-3}$.  

Fortunately, there is a clean way to see that the DM cannot contain a sufficient density of $ \bar{H}$'s to account for the BAU.  When an $\bar{X}$ annihilates, an energy $m_X + B_X m_N$ is released, almost all of which is converted to heat.  Uranus provides a remarkable system for constraining such a possibility, because of its large size and extremely low level of heat production, $42 \pm 47 \, {\rm erg ~ cm}^{-2} s^{-1}$ \cite{uranusVoyager}.  When annihilation is in equilibrium with capture as discussed above, the heat flux supplied by annihilation is
$ f_{\rm cap}^U  \Phi_{\bar{X}} (m_X + B_X m_N).$
For the $\bar{H}$, $f_{\rm cap}^U \approx 0.2$ as for Earth, so the heat flux generated in Uranus should be $470  \, {\rm erg ~ cm}^{-2}s^{-1}$, which definitively excludes the $H,\,\bar{H}$ scenario. However the $\bar{X_4}$ scenario is safe by many orders of magnitude, because $ f_{\rm cap}^U  = 3 \cdot 10^{-5}$ with the ``conservative" DM halo velocity distribution\cite{zf:window}, giving a heat flux of $0.07  \, {\rm erg ~ cm}^{-2}s^{-1}$ or even less if equilibrium is not established.  

We now turn to the particle physics of a new, light fundamental particle with $B_X = 1$ and $m_X \lsi 4.5$ GeV.  Such a low mass suggests it is a fermion whose mass is protected by a chiral symmetry.  Various dimension-6 interactions with quarks could appear in the low energy effective theory after high scale interactions, e.g., those responsible for the family structure of the Standard Model, have been integrated out.  These include
\begin{equation} \label{Xbcd}
\frac{g}{M^2} (\bar{X} b \, \bar{d^c} c - \bar{X} c \, \bar{d^c} b) + h.c.,
\end{equation}
where the
$b$ and $c$ fields are left-handed SU(2) doublets combined to form an SU(2) singlet and $d^c$ is the charge conjugate of the SU(2) singlet field $b_R$.  The suppressed color and spin indices are those of the antisymmetric operator $\tilde{O}^{\dot{a}}$ given in equation (10) of ref. \cite{peskin79}.  The hypercharge of the left-handed quarks is +1/3 and that of $d_R$ is -2/3, so the $X$ is a singlet under all standard model interactions and its only interaction with fields of the standard model are through operators such as (\ref{Xbcd}).  Dimension-6 operators involving only third generation quarks can be constructed; supplemented by $W$ exchange or penguins, they could also be relevant.  

Prior to freezeout, $\bar{X}$'s stay in equilibrium through scattering reactions like 
\begin{equation}
\label{dXbar}
d + \bar{X} \leftrightarrow \bar{b}~\bar{c}. 
\end{equation}
The coupling $g$ in (\ref{Xbcd}) is in general complex and a variety of diagrams involving all three generations and including both W exhange and penquins contribute to generating the effective interaction in (\ref{dXbar}), so the conditions \newpage \noindent necessary for a sizeable CP-violating asymmetry between $\sigma_{X}^{ \rm ann} $ 
and $  
\sigma_{\bar{X}}^{ \rm ann}$ are in place.  An interaction such as (\ref{Xbcd}) gives rise to $ \sigma_{\bar{X} d \rightarrow \bar{b} \bar{c}} \sim (g/M^2)^2 m_X T_{\rm f.o.}.$  Therefore $g/M^2 \approx 10^{-9} - 10^{-10}\, {\rm GeV}^{-2}$ is needed for freezeout to occur at the correct temperature.  It is not possible to use this estimate of $(g/M^2)$ to obtain $\sigma^{\rm el}_{XN \{\bar{X} N\} }$ without understanding how the interaction (\ref{Xbcd}) is generated, since it is not renormalizable.  A naive guess $\sigma^{\rm el}_{XN \{\bar{X} N\} } \sim (g/M^2)^4 m_X^6 \,10 \, {\rm mb} \approx 10^{-40} \,{\rm to} \, 10^{-43} {\rm cm}^2$ is well below the present limit of $\approx 10^{-38} {\rm cm}^2$ for a 4 GeV particle, but the actual value depends on the high-scale physics and could be significantly larger or smaller.  

If (\ref{Xbcd}) is the only coupling of $X$ to quarks,  the effects of annihilation in Earth, Sun, Uranus and galactic center are unobservably small.  This is because $m_X + m_N < m_B + m_D$, so an additional factor $\approx G_F^2 |V_{bc}|^2 {\rm GeV}^4$ is required below the chiral phase transition leading to $\sigma^{\rm ann}_{\bar{X}N}  \approx 5 \times 10^{-58} {\rm cm}^2$.  However there could be other interactions which are subdominant at freezeout but are kinematically allowed in the low temperature phase, e.g., (\ref{Xbcd}) with $b \rightarrow d$, so a search for an unexpected neutrino or gamma-ray flux is worthwhile, even though null results would not exclude the model.

To summarize, we have shown that the dark matter and baryon asymmetry puzzles may be related.  We presented two concrete scenarios, in which the observed values of $\Omega_{DM}$ and $\Omega_b$ are explained.  One of them entails a long-lived H dibaryon, but it is excluded by limits on heat production in Uranus.  The other involves a new  $\sim 4$ GeV particle with dimension-6 couplings to quarks.  This scenario can arise naturally from beyond-the-standard-model particle physics, but it suggests that the DM particle may be elusive. \\

\noindent{\bf Acknowledgements: } 
This research was supported in part by NSF-PHY-0101738.  We have benefited from discussions with many colleagues; special thanks go to S. Mitra for bringing the strong limits on heating in Uranus to our attention, and to S. Nussinov for discussions about ways to exclude the H dibaryon scenario, in particular the idea of using white dwarf heating which although unsuccessful planted the idea we used here with Uranus. GRF also thanks the Departments of Physics and Astronomy of Princeton University for their support and hospitality during the course of this research.


\end{document}